# Optimal firm's policy under lead time- and price-dependent demand: interest of customers rejection policy

Abduh Sayid Albana
Université Grenoble Alpes, G-SCOP, F-38000 Grenoble, France
abduh-sayid.albana@grenoble-inp.org
Yannick Frein
Université Grenoble Alpes, G-SCOP, F-38000 Grenoble, France
Ramzi Hammami
ESC Rennes School of Business, 35065 Rennes, France

**Abstract**

Considering a lead-time- and price-sensitive demand, we investigate whether a client rejection policy, modeled as M/M/1/K system, can be more profitable than an all-client acceptance policy, modeled as M/M/1 system. We provide analytical insights for the cases with and without holding and penalty costs by comparing M/M/1/1 to M/M/1 models.

## INTRODUCTION

The delivery lead-time, which represents the elapsed time between the placement of the order by the customer and the receipt of this order, has become a factor of competitiveness for companies and an important purchase criterion for many customers. Geary and Zonnenberg (2000) reported that top performers among 110 organizations conducted initiatives not only to reduce costs and maintain reliability, but also to improve delivery speed and flexibility. Baker et al. (2001) found that less than 10% of end consumers and less than 30% of corporate customers base their purchasing decisions on price only; for a substantial majority of purchasers both price and delivery lead time are crucial factors that determine their purchase decisions. Thus, in order to increase their profit, companies must not focus only on price but also need to quote the right delivery lead time to their customers. A short quoted lead time can lead to higher demand but can also result in late delivery, which affects the firm's reputation for on-time delivery and deters future customers (Slotnick, 2014). In addition, companies risk to lose markets if they are not capable of respecting the quoted delivery lead-times (Kapuscinski and Tayur, 2007). A long quoted lead time can reduce the risk of late delivery but leads to lower demand. This raises the following relevant question: What is the best lead time that must be quoted by a company when customers are not only sensitive to price but also to lead time? Some authors tried to answer this question while considering an M/M/1 system (Palaka et al., 1998; Pekgün et al., 2008; So and Song, 1998). One of the characteristics of the M/M/1 is accepting all customers, which can lead to congestions in the queue.

In order to cope with this situation, firms can react by quoting longer lead times in order to maintain the desired service level. However, this leads to lower demand and revenue. Starting from this observation, we investigate in this paper whether a customer rejection policy can be more beneficial for the firm than an all-customers acceptance policy. Indeed, our idea is based on the fact that rejecting some customers might help to quote shorter lead time for the accepted customers, which might lead to higher demand and profit. We model this rejection policy based on an M/M/1/K system. We analytically determine the optimal firm's policy (optimal price and quoted lead time) in case of M/M/1/1 system. Then, we compare the optimal firm's profit under M/M/1/1 with the optimal profit obtained by M/M/1. Two situations are considered: a system without holding and penalty costs and a system where these costs are included.

The rest of this paper is organized as follows. In the next section, we develop the formulation of the M/M/1/K system with price-and lead-time-sensitive demand. Later, we analytically solve the M/M/1/K system for K=1 without holding and penalty cost and compare the results to those obtained with M/M/1. We then focus on the case with holding and penalty cost for the rest of the paper. We finally conclude and give future work directions.

## PROPOSED MODEL (THE M/M/1/K)

As in Palaka et al. (1998), we consider a make-to-order firm where the capacity is assumed to be constant while price, quoted lead-time, and, consequently, demand are decision variables. Customers are served in first-come, first-served basis (FCFS). The arrival process is assumed to be a Poisson process. The processing times of customers in the system are assumed to be exponentially distributed. Contrary to the assumptions of M/M/1 model where all customers are accepted (as in Palaka et al., 1998 and Pekgun et al., 2008), we reject customers when we have already K customers in the system, therefore we model the system by an M/M/1/K.

Similarly to Liu et al. (2007); Palaka et al. (1998); and Pekgün et al., (2008), the demand is assumed to be a linear decreasing function in price and quoted lead-time.

$$\Lambda(p,l) = a - b_1 p - b_2 l, \qquad (1)$$

where:
- $p$ = price of the good/service set by the firm,
- $l$ = quoted lead-time,
- $\Lambda(p,l)$ = expected demand for the good/service with price $p$ and quoted lead-time $l$,
- $a$ = market potential,
- $b_1$ = price sensitivity of demand,
- $b_2$ = lead-time sensitivity of demand,

Since the demand is downward sloping in both price and quoted lead-time, $b_1$ and $b_2$ are restricted to be non-negative

According to Palaka et al. (1998), this linear demand function is tractable and has several desirable properties. For instance, with such a linear demand, the price elasticity is increasing in both price and quoted lead-time. Customers would be more sensitive to long lead-times when they are paying more for the goods or service. Similarly, customers would be more sensitive to high prices when they also have longer waiting times.

In order to prevent the firms from quoting unrealistically short lead-times, we assume that the firm maintains a certain minimum service level. The service level ($s$) is defined as the probability that the actual lead-time ($W$) satisfies the quoted lead-time ($l$) ($P(W \leq l) \geq s$).

Since we assume an M/M/1/K queueing system with mean service rate, $\mu$, mean arrival rate (or demand), $\lambda$, and throughput rate (effective demand), $\bar{\lambda}$, then the expected number of customers in the system is given by $L_s$ (see eq. (2)), and the actual lead-time (time in the system) is exponentially distributed with mean $L_s/\bar{\lambda}$ (see eq. (3)). The probability that the firm is able to meet the quoted lead-time ($P(W \leq l)$) and the probability that a job is late ($P(W > l)$) are given in eq. (4). These equations are based on Gross et al. (2008).

$$L_s = \frac{\rho}{1-\rho} - \frac{(K+1)\rho^{K+1}}{1-\rho^{K+1}} \text{ with } \rho = \frac{\lambda}{\mu} \tag{2}$$

$$W = \frac{L_s}{\bar{\lambda}} \text{ with } \bar{\lambda} = \lambda(1-P_K) \text{ and } P_K = \frac{1-\rho}{1-\rho^{K+1}}\rho^K \text{ if } \rho \neq 1 \text{ or } P_K = \frac{1}{K+1} \text{ if } \rho = 1 \tag{3}$$

$$P(W \leq l) = 1 - \sum_{k=0}^{K-1}\left(\sum_{i=0}^{k}\frac{(\mu \cdot l)^i}{i!}e^{-\mu \cdot l}\right)\frac{P_k}{1-P_K} \text{ and } P(W > l) = \sum_{k=0}^{K-1}\left(\sum_{i=0}^{k}\frac{(\mu \cdot l)^i}{i!}e^{-\mu \cdot l}\right)\frac{P_k}{1-P_K} \tag{4}$$

The objective of the firm is to maximize its revenue, which includes the following three parts:
   (1) **Expected revenue (net of direct costs)** is represented by $\bar{\lambda}(p-m)$, where $m$ is the unit direct variable cost.
   (2) **Total Congestion cost** is expressed as the mean number of jobs in the system multiplied by the unit holding cost ($L_s \times F$). This cost typically represents the in-process inventory holding cost.
   (3) **Total Lateness penalty cost** is expressed as (penalty per job per unit lateness) × (number of overdue clients) × (expected lateness given that a job is late). The number of overdue clients is equal to: (throughput rate) × (probability that a job is late). The penalty cost per job per unit lateness (denoted by $c$) reflects the direct compensation paid to customers for not meeting the quoted lead-time. Mathematically, this total Lateness penalty cost is given by $(c \times \bar{\lambda} \times P(W \geq l) \times W)$.

Thus, the firm's optimization problem can be formulated as follows:

(P0) Maximize $\bar{\lambda}(p-m) - (L_s \times F) - (c \times \bar{\lambda} \times P(W \geq l) \times W)$ (5)
$\quad l, p, \lambda$

Subject to $\lambda \leq a - b_1 p - b_2 l$ (6)

$$1 - \sum_{k=0}^{K-1}\left(\sum_{i=0}^{k}\frac{(\mu \cdot l)^i}{i!}e^{-\mu \cdot l}\right)\frac{P_k}{1-P_K} \geq s \tag{7}$$

$$\rho = \frac{\lambda}{\mu} \tag{8}$$

$$P_K = \frac{1-\rho}{1-\rho^{K+1}}\rho^K \text{ if } \rho \neq 1 \text{ and } P_K = \frac{1}{K+1} \text{ if } \rho = 1 \tag{9}$$

$$\bar{\lambda} = \lambda(1-P_K) \tag{10}$$

$$\lambda, p, l, \bar{\lambda} \geq 0 \tag{11}$$

where,

| Decision Variables | | Parameters | |
|---|---|---|---|
| $p$ = Price of the good/service set by the firm, | | $a$ = Market potential | |
| $l$ = Quoted lead-time, | | $b_1$ = Price sensitivity of demand, | |
| $\lambda$ = Mean arrival rate (demand), | | $b_2$ = Lead-time sensitivity of demand, | |
| | | $\mu$ = Mean service rate (Production capacity), | |
| | | $m$ = Unit direct variable cost, | |
| | | $s$ = Service level defined by company, | |
| | | $P_K$ = Probability of rejected customer, | |
| | | $K$ = System capacity. | |

In this formulation, constraint (6) imposes that the mean demand ($\lambda$) does not exceed the demand obtained with price ($p$) and quoted lead-time ($l$). Constraint (7) expresses the service level constraint. Constraint (9) calculates the probability of rejecting customers. Constraint (10) is the number of customers that are served and exit the system. Constraints (11) are the non-negativity constraints.

Solving analytically this general case seems to be difficult. So, in the following section, we only consider the case of K = 1. We will consider two situations: the case without penalty and holding cost; and the case where these costs are included. In both cases, we will compare the obtained optimal solution with the optimal solution of the M/M/1 approach and derive insights.

## M/M/1/1 WITHOUT PENALTY COST AND HOLDING COST

We consider here the case without holding and penalty costs. Thus, the objective function consists only in the maximization of the expected revenue. The Probability $P(W > l) = e^{-\mu l}$, hence implying that the service level constraint can be written as $1 - e^{-\mu l} \geq s$. Consequently, the formulation of the problem becomes:

(P1) $\quad \underset{\lambda, l, p, \bar{\lambda}}{\text{Maximize}} \quad \bar{\lambda}(p - m)$ \hfill (12)

$\quad$ Subject to $\quad \lambda \leq a - b_1 p - b_2 l$ \hfill (13)

$\qquad\qquad\qquad 1 - e^{-\mu l} \geq s$ \hfill (14)

$\qquad\qquad\qquad \rho = \lambda / \mu$ \hfill (15)

$\qquad\qquad\qquad P_1 = \dfrac{\rho}{1 + \rho}$ \hfill (16)

$\qquad\qquad\qquad \bar{\lambda} = \lambda(1 - P_1)$ \hfill (17)

$\qquad\qquad\qquad \lambda, p, l, \bar{\lambda} \geq 0$ \hfill (18)

Eq. (14) can be rewritten as $\mu l \geq \ln(1/(1-s))$. Then, by integrating the equality constraint (eq. (15-17)) into the objective function, we can simplify the formulation as:

(P1') Maximize $_{\lambda,l,p}$ $(\lambda\mu/(\mu+\lambda))\times(p-m)$ (19)

Subject to $\lambda \leq a - b_1 p - b_2 l$ (20)

$\mu l \geq \ln(1/(1-s))$ (21)

$\lambda, p, l \geq 0$ (22)

Using the new formulation (P1') we will now transform the problem into a single variable optimization problem. Because of the space limitation, all the proofs are reported in Albana et al. (2016). Firstly, the demand constraint (eq. (20)) is binding at optimality, thus: $\lambda = a - b_1 p - b_2 l$ $\Leftrightarrow p = (a - b_2 l - \lambda)/b_1$ (Albana et al., 2016). Secondly, the service level constraint (eq. (21)) is also binding at optimality (Albana et al., 2016). Hence, $l = \ln(1/(1-s))/\mu$. We denote $\ln(1/(1-s))$ by $z$, and get: $l = z/\mu$. Substitute $l = z/\mu$ into the expression of $p$, we obtain: $p = (a\mu - b_2 z - \lambda\mu)/\mu b_1$. Substituting $p$ into the objective function, we get a new formulation of the problem with single variable ($\lambda$) as:

(P1'') Maximize $_{\lambda \geq 0}$ $f(\lambda) = \left(\dfrac{\lambda a\mu - \lambda b_2 z - \lambda^2 \mu - \lambda m\mu b_1}{\mu b_1 + \lambda b_1}\right)$ (23)

**Proposition 1.** There exists $\lambda \geq 0$ such as $f(\lambda) \geq 0$ iff $\dfrac{a\mu - b_2 z}{\mu b_1} \geq m$ (proof see Albana et al. (2016)).

In order to find the optimal solution, we use the first derivative conditions. Indeed,

$$\dfrac{df(\lambda)}{d\lambda} = \dfrac{d}{d\lambda}\left(\dfrac{\lambda a\mu - \lambda b_2 z - \lambda^2 \mu - \lambda m\mu b_1}{\mu b_1 + \lambda b_1}\right) = 0 \Leftrightarrow \dfrac{\mu b_1 (a\mu - b_2 z - m\mu b_1 - 2\lambda\mu - \lambda^2)}{(\mu b_1 + \lambda b_1)^2} = 0 \quad (24)$$

The numerator of eq. (24) must be equal to zero. It can be proven that the discriminant ($\Delta$) of this numerator under condition of proposition 1 is greater or equal to 0. Hence eq. (24) has two real roots:

$$\lambda_1 = -\mu - \sqrt{\mu^2 + a\mu - b_2 z - m\mu b_1} \text{ and } \lambda_2 = -\mu + \sqrt{\mu^2 + a\mu - b_2 z - m\mu b_1} \quad (25)$$

Given that $\lambda_1$ is negative, there is only one feasible stationary point $\lambda_2$. Under proposition 1, $\lambda_2$ is positive. It can also be shown that the objective function is a concave function in $\lambda, l, p \geq 0$ (Albana et al., 2016). Hence, $\lambda_2$ is also the optimum point. This leads to the results presented in proposition 2.

**Proposition 2.** For problem P1, the optimum demand is $\lambda^* = -\mu + \sqrt{\mu^2 + a\mu - b_2 z - m\mu b_1}$ with $z = \ln(1/(1-s))$, the optimum lead-time $l^* = \ln(1/(1-s))/\mu$, the optimum price $p^* = (a - b_2 l^* - \lambda^*)/b_1$, and the optimum profit = $(\lambda^*\mu/(\mu+\lambda^*)) \times (p^* - m)$.

## COMPARISON BETWEEN M/M/1/1 AND M/M/1 WITHOUT HOLDING AND PENALTY COST

In this section, we compare our model (M/M/1/1) with the existing M/M/1 taken from Pekgün et al. (2008) as they don't consider holding and penalty costs. We use a base case with parameters: $b_1 = 4$; $\mu = 10$; $s = 0.95$; $m = 5$. We vary the market potential ($a$) and the lead-time sensitivity ($b_2$). For each pair of value ($a$, $b_2$), we calculate the relative gain obtained by using M/M/1/1 instead of M/M/1. This relative gain is calculated as follows:

$$\frac{\text{Profit}^{M/M/1/1} - \text{Profit}^{M/M/1}}{\text{Profit}^{M/M/1}} \times 100\% \quad (26)$$

A positive value means that the approach with rejections (M/M/1/1 model) is better than the approach without rejections (M/M/1 model).

*Table 1 - Comparison for different values of a and $b_2$*

| $b_2$ | M/M/1 vs M/M/1/1 | | | | |
|---|---|---|---|---|---|
| 20 | 40.87% | 17.94% | 8.29% | 3.42% | 0.66% |
| 19 | 37.10% | 15.27% | 6.12% | 1.57% | -0.99% |
| 18 | 33.39% | 12.61% | 3.96% | -0.30% | -2.65% |
| 17 | 29.73% | 9.96% | 1.79% | -2.18% | -4.33% |
| 16 | 26.13% | 7.31% | -0.40% | -4.07% | -6.03% |
| 15 | 22.57% | 4.66% | -2.59% | -5.99% | -7.74% |
| 14 | 19.05% | 2.01% | -4.80% | -7.92% | -9.48% |
| 13 | 15.57% | -0.65% | -7.03% | -9.87% | -11.24% |
| 12 | 12.11% | -3.32% | -9.29% | -11.86% | -13.04% |
| 11 | 8.68% | -6.02% | -11.57% | -13.88% | -14.87% |
| 10 | 5.26% | -8.74% | -13.90% | -15.94% | -16.74% |
| 9 | 1.85% | -11.49% | -16.28% | -18.06% | -18.67% |
| 8 | -1.56% | -14.30% | -18.71% | -20.23% | -20.65% |
| 7 | -4.98% | -17.17% | -21.22% | -22.48% | -22.71% |
| 6 | -8.43% | -20.13% | -23.83% | -24.83% | -24.86% |
| 5 | -11.92% | -23.20% | -26.56% | -27.30% | -27.14% |
| | 30 | 40 | 50 | 60 | 70 |
| | $a$ | | | | |

We can see in table 1 that the rejection policy can be better than the all-customers acceptance policy, in particular when the market potential is small and the lead-time sensitivity is high.

Thus, the client rejection policy can be better in some cases even when the penalty and holding costs are removed. This motivates the purpose of next section where we will study the case with penalty and holding costs

## M/M/1/1 WITH PENALTY COST AND HOLDING COST

With the addition of penalty and holding costs, the objective function includes the expected revenue, the total congestion costs, and the total lateness penalty costs. The formulation of this objective function has been presented earlier. The service level constraint is similar to the previous case. Thus, the formulation of the problem is:

$$(P2) \quad \underset{\lambda, p, l, \bar{\lambda}}{\text{Maximize}} \quad \bar{\lambda}(p-m) - \frac{F\lambda}{\mu+\lambda} - \frac{c\bar{\lambda}}{\mu} e^{-\mu l} \tag{27}$$

$$\text{Subject to} \quad \lambda \leq a - b_1 p - b_2 l \tag{28}$$

$$1 - e^{-\mu l} \geq s \tag{29}$$

$$\rho = \lambda/\mu \tag{30}$$

$$P_1 = \frac{\rho}{1+\rho} \tag{31}$$

$$\bar{\lambda} = \lambda(1 - P_1) \tag{32}$$

$$\bar{\lambda}, \lambda, l, p \geq 0 \tag{33}$$

Integrating the equality constraints (eq. (30 – 32)) to the objective function and rewriting $1 - e^{-\mu l} \geq s$ as $\mu l \geq \ln(1/(1-s))$, we get the following formulation of the problem:

$$(P2') \quad \underset{\lambda, l, p}{\text{Maximize}} \quad \frac{\lambda(p\mu - m\mu - F - ce^{-\mu l})}{\mu + \lambda} \tag{34}$$

$$\text{Subject to} \quad \lambda \leq a - b_1 p - b_2 l \tag{35}$$

$$\mu l \geq \ln(1/(1-s)) \tag{36}$$

$$\lambda, p, l \geq 0 \tag{37}$$

It can be shown that the demand constraint (eq. (35)) is binding at optimality (Albana et al., 2016). Thus, the formulation becomes:

$$(P2'') \quad \underset{\lambda, l}{\text{Maximize}} \quad \frac{\lambda\left[(\mu(a - b_2 l - \lambda)/b_1) - m\mu - F - ce^{-\mu l}\right]}{\mu + \lambda} \tag{38}$$

$$\text{Subject to} \quad \mu l \geq \ln(1/(1-s)) \tag{39}$$

$$l, p \geq 0 \tag{40}$$

In addition, there is also a feasibility condition of this problem as explained in proposition 3.

**Proposition 3.** The problem P2" is feasible ($\lambda, l \geq 0$ and profit $\geq 0$) iff $\dfrac{a - b_2 l}{b_1} \geq m$ and $a\mu - \mu b_2 l - \mu m b_1 - F b_1 - c b_1 e^{-\mu l} \geq 0$ (proof see Albana et al. (2016)).

Unlike the first case where the penalty and holding costs are not considered, the service level constraint is not necessarily binding in this case, which complicates the solving approach. Indeed, for large values of $c$, the actual service level has to be very high (close to 1) to avoid a high penalty cost. This indicates that the actual service level can be greater than the imposed service level ($s$). We now present the main steps to get the optimal solution given in proposition 4. The detailed proof is given in (Albana et al., 2016).

To solve the problem, we apply the Lagrangian multiplier method. The stationary points of problem (P2") must satisfy:

$$\frac{\partial L}{\partial \lambda} = 0, \frac{\partial L}{\partial l} = 0, \gamma \frac{\partial L}{\partial \gamma} = 0, \gamma \geq 0 \text{ and } \mu l \geq \ln\left(\frac{1}{1-s}\right)$$

where,

$$L(\lambda, l, \gamma) = \frac{\lambda\left[\left(\mu(a - b_2 l - \lambda)/b_1\right) - m\mu - F - ce^{-\mu l}\right]}{\mu + \lambda} + \gamma\left\{\mu l - \ln\left(\frac{1}{1-s}\right)\right\} \quad (41)$$

As we have already explained, we have two situations:
  (1) the service level constraint (39) is non-binding: $s < s_c$,
  (2) the service level constraint (39) is binding: $s \geq s_c$,

where the critical value for the service level ($s_c$) equals to $1 - b_2/b_1 c$. With this critical service level ($s_c$) we can prove that the actual service level is equal to $Max\{s, s_c\}$ (Albana et al., 2016).

$$1 - e^{-\mu l} = Max\{s_c, s\} \Leftrightarrow l^* = \frac{1}{\mu}\ln\left(\frac{1}{1 - Max\{s_c, s\}}\right)$$

$$\Leftrightarrow l^* = \frac{1}{\mu}\ln x \text{ where, } x = Max\{1/(1-s), b_1 c/b_2\} \quad (42)$$

To find the optimal demand, we derive eq. (41) by $\lambda$.

$$\frac{\partial L}{\partial \lambda} = 0 \Leftrightarrow \left[\mu\left(a\mu - \mu b_2 l - 2\mu\lambda - \mu m b_1 - F b_1 - b_1 c e^{-\mu l} - \lambda^2\right)\right]/\left[b_1(\mu + \lambda)^2\right] = 0 \quad (43)$$

The numerator of eq. (43) should be equal to zero. It is proven that, with the feasibility condition, the discriminant ($\Delta$) of eq. (43) is positive (Albana et al., 2016). Hence eq. (43) has two roots which are:

$$\lambda_1 = -\mu - \sqrt{\mu^2 + a\mu - \mu b_2 l - \mu b_1 m - F b_1 - b_1 c e^{-\mu l}} \quad \text{and}$$

$$\lambda_2 = -\mu + \sqrt{\mu^2 + a\mu - \mu b_2 l - \mu b_1 m - F b_1 - b_1 c e^{-\mu l}} \tag{44}$$

$\lambda_1$ is negative. Under feasibility condition of proposition 3, $\lambda_2$ is positive. It is proven that the objective is concave function for $\lambda, l, p \geq 0$ (Albana et al., 2016). Thus, the lead-time ($l^*$) and demand ($\lambda_2$) provide the optimal solution. As a summary, the optimum point of this problem is given in the proposition 4.

---

**Proposition 4.** For problem P2:
1. The optimum lead-time $l^* = \ln x / \mu$ with $x = Max\{1/(1-s), b_1 c / b_2\}$.
2. The optimum demand can be found by using equation $\lambda^* = -\mu + \sqrt{\mu^2 + a\mu - \mu b_2 l^* - \mu b_1 m - F b_1 - b_1 c e^{-\mu l^*}}$, the optimum price $p^* = (a - b_2 l^* - \lambda^*)/b_1$.
3. The optimum profit $= \lambda^* (p^* \mu - m\mu - F - c e^{-\mu l^*})/(\mu + \lambda^*)$.

---

Based on the result found in this section, we will investigate whether our M/M/1/1 model with penalty and holding costs can be better than the M/M/1 model of Palaka et al. (1998).

## COMPARISON BETWEEN M/M/1/1 AND M/M/1 WITH HOLDING AND PENALTY COST

In this section, we compare our model (M/M/1/1) with the existing M/M/1 taken from Palaka et al. (1998). We vary the market potential (*a*) and the lead-time sensitivity (*b₂*) (see Table 2). For each pair of value (*a*, *b₂*), we calculate the relative gain obtained by using M/M/1/1 instead of M/M/1. This relative value follows equation (26). We use the same base case as in the previous comparison with additions of $F = 2$ and $c = 10$.

As expected, we have more cases where M/M/1/1 is better than M/M/1 in this situation with penalty and holding costs compared to the situation without holding and penalty costs. In the M/M/1, because all clients are accepted, the lead-time can be very long which can cause high congestion costs. As $b_2$ increases, the demand becomes more sensitive to lead-time, which favors the rejection policy modeled by the M/M/1/1.

## CONCLUSION

We have provided the general formulation of M/M/1/K with penalty and holding costs where demand is sensitive to both price and lead-time. We solved analytically the case with K=1 in two situations: with and without penalty and holding costs. We conducted numerical experiments based on the analytical solutions and showed that the rejection policy (M/M/1/1) can be better than the all-customers acceptance policy (M/M/1). This paper can be extended in many ways such as considering the case of K>1 or modeling a system of type M/D/1. We are currently working on these issues.

*Table 2 - Comparison for different values of a and $b_2$*

| $b_2$ | M/M/1 vs M/M/1/1 | | | | |
|---|---|---|---|---|---|
| 20 | 53.96% | 26.95% | 15.50% | 9.58% | 6.11% |
| 19 | 49.95% | 24.23% | 13.33% | 7.74% | 4.49% |
| 18 | 46.02% | 21.53% | 11.17% | 5.90% | 2.86% |
| 17 | 42.16% | 18.84% | 9.02% | 4.05% | 1.22% |
| 16 | 38.37% | 16.17% | 6.86% | 2.19% | -0.43% |
| 15 | 34.64% | 13.51% | 4.69% | 0.33% | -2.09% |
| 14 | 30.96% | 10.85% | 2.52% | -1.54% | -3.76% |
| 13 | 27.34% | 8.20% | 0.34% | -3.43% | -5.45% |
| 12 | 23.77% | 5.56% | -1.85% | -5.34% | -7.16% |
| 11 | 20.24% | 2.91% | -4.05% | -7.26% | -8.89% |
| 10 | 16.74% | 0.25% | -6.27% | -9.21% | -10.65% |
| 9 | 13.28% | -2.42% | -8.52% | -11.19% | -12.43% |
| 8 | 9.84% | -5.10% | -10.80% | -13.19% | -14.25% |
| 7 | 6.42% | -7.81% | -13.11% | -15.24% | -16.10% |
| 6 | 3.01% | -10.56% | -15.47% | -17.33% | -18.01% |
| 5 | -0.40% | -13.35% | -17.88% | -19.49% | -19.97% |
| | 30 | 40 | 50 | 60 | 70 |
| | *a* | | | | |